\documentclass[12pt]{article}
\usepackage[cmtip,arrow]{xy}
\usepackage{pb-diagram,pb-xy}
\usepackage{amsmath}
\usepackage[psamsfonts]{amssymb}
\usepackage{cmmib57}
\usepackage{amsmath,amssymb,amscd}
\newcommand{\bPf}{\par\vspace*{-4pt}\indent{\sc Proof.}\enskip}
\newcommand{\ePf}{\medskip}
\def\QED{\hskip0.1em\hfill\null\ \null\nobreak\hfill\kern3pt\vbox{\hrule\hbox
   {\vrule\kern1pt\vbox{\kern1.7pt\hbox{$\scriptscriptstyle{QED}$}
    \kern0.2pt}\kern1pt\vrule}\hrule}}

\def\END{\hskip0.1em\hfill\null\ \null\nobreak\hfill\kern3pt\vbox{\hrule\hbox
   {\vrule\kern1pt\vbox{\kern1.7pt\hbox{$\,\,\,\vspace{5pt}$}
    \kern0.2pt}\kern1pt\vrule}\hrule}}
\newtheorem{theorem}{Theorem}
\newtheorem{lemma}{Lemma}
\newtheorem{corollary}{Corollary}
\newtheorem{proposition}{Proposition}
\newtheorem{remark}{Remark}
\newtheorem{definition}{Definition}
\newtheorem{example}{Example}
\newcommand{\bCd}{\bEq\begin{CD}}
\newcommand{\eCd}{\end{CD}\eEq}
\newcommand{\bcd}{\beq\begin{CD}}
\newcommand{\ecd}{\end{CD}\eeq}
\newcommand{\ben}{\begin{enumerate}}
\newcommand{\een}{\end{enumerate}}
\newcommand{\bEq}{\begin{eqnarray}}
\newcommand{\eEq}{\end{eqnarray}}
\newcommand{\beq}{\begin{eqnarray*}}
\newcommand{\eeq}{\end{eqnarray*}}
\newcommand{\bDf}{\begin{definition}\em}
\newcommand{\eDf}{\end{definition}}
\newcommand{\bLm}{\begin{lemma}}
\newcommand{\eLm}{\end{lemma}}
\newcommand{\bPr}{\begin{proposition}}
\newcommand{\ePr}{\end{proposition}}
\newcommand{\bTh}{\begin{theorem}}
\newcommand{\eTh}{\end{theorem}}
\newcommand{\bCr}{\begin{corollary}}
\newcommand{\eCr}{\end{corollary}}
\newcommand{\bRm}{\begin{remark}\em}
\newcommand{\eRm}{\end{remark}}
\newcommand{\bEx}{\begin{example}\em}
\newcommand{\eEx}{\end{example}}

\newcommand{\ie}{{\em i.e$.$} }
\newcommand{\eg}{{\em e.g$.$} }
\newcommand{\R}{I\!\!R}

\newcommand{\mto}{\mapsto}

\newcommand{\cE}{\mathcal{E}}

\newcommand{\cI}{\mathcal{I}}

\newcommand{\cL}{\mathcal{L}}

\newcommand{\cR}{\mathcal{R}}

\newcommand{\bI}{\boldsymbol{I}}
\newcommand{\bJ}{\boldsymbol{J}}

\newcommand{\bX}{\boldsymbol{X}}
\newcommand{\bY}{\boldsymbol{Y}}

\newcommand{\wed}{\wedge}
\newcommand{\com}{\!\circ\!}

\newcommand{\alp}{\alpha}
\newcommand{\bet}{\beta}
\newcommand{\gam}{\gamma}
\newcommand{\del}{\delta}
\newcommand{\eps}{\epsilon}
\newcommand{\zet}{\zeta}

\newcommand{\iot}{\iota}

\newcommand{\lam}{\lambda}

\newcommand{\ome}{\omega}

\newcommand{\Lam}{\Lambda}

\newcommand{\For}{{\Lambda}}

\newcommand{\Var}{{\mathcal{V}}}
\newcommand{\Thd}{{\Theta}}
\title{\large{\bf
Variational derivatives in locally Lagrangian field theories and Noether--Bessel-Hagen currents
}}
\author{{\normalsize F. Cattafi}
\\
{\footnotesize Department of Mathematics, Universiteit Utrecht}
\\
{\footnotesize 3508 TA Utrecht, The Netherlands}
\\  
{\footnotesize e--mail: 
{\sc  f.cattafi@uu.nl
}}
\\
{\normalsize M.
Palese
\, and E. Winterroth
}
\\ {\footnotesize Department of Mathematics,
University of Torino}
\\
{\footnotesize via C. Alberto 10, 10123 Torino, Italy} 
\\  {\footnotesize e--mail: 
{\sc marcella.palese@unito.it, ekkehart.winterroth@unito.it
}}
}%add addresses
\begin{document}
\maketitle

\begin{abstract}

The variational Lie derivative of classes of forms in the Krupka's variational sequence is defined as a variational Cartan formula at any degree, in particular  for degrees lesser than the dimension of the basis manifold. As an example of application we determine the condition for a Noether--Bessel-Hagen current, associated with a generalized symmetry, to be variationally equivalent to a Noether current for an invariant Lagrangian. We show that, if it exists, this Noether current is exact on-shell and generates a canonical conserved quantity.

\end{abstract}


%-----------------------------------------------------------------------------%
% T I T L E
%-----------------------------------------------------------------------------%

%-------------------------------------------------------------------------%
\section{Introduction} 
%-------------------------------------------------------------------------%

The study of calculus of variations for field theories (multiple integrals) as a theory of differential forms and their exterior differential modulo contact forms (`congruences') was initiated \cite{Lep36} by Lepage in 1936; see \eg \cite{MuLe09} for a brief review. 
One of the most important fact within such a geometric formulation
of the calculus of variations is the fact that considering the
ambient manifold to be a fibered manifold and the configuration
space a jet prolongation of it, {\em variations can be described by Lie
derivatives of forms with respect to projectable vector fields};
see, \eg \cite{GoSt73,Kru73}.

In \cite{Kru90, Kru95} Krupka described the contact structure at a
given finite prolongation order and initiated the project of
framing the calculus of variations within a differential sequence
obtained as a quotient sequence of the de Rham sequence. Krupka's
variational sequence is a sequence of differential forms modulo a contact structure inspired by the Lepage idea of a `congruence'. Krupka also showed that the Lie derivative of forms with
respect to projectable vector fields preserves  the contact
structure naturally induced by the affine bundle structure of jet
projections order-by-order.
This fact suggests that a Lie derivative of classes of forms, \ie a {\em variational Lie derivative}, can be correctly defined as the equivalence class of the standard Lie derivative of forms and represented by forms.

By a representation of the quotient sheaves of the variational sequence
as sheaves of sections of tensor bundles, in \cite{FrPaVi02} explicit formulae were
provided for the quotient Lie derivative operators, as well as
corresponding versions of Noether Theorems interpreted in terms of
conserved currents for Lagrangians and Euler--Lagrange morphisms (only classes of forms up to degree $n+2$,
the latter assumed to be exact, were considered).
Such a representation made use of intrinsic decomposition formulae
for  vertical morphisms due to Kola\v r \cite{Kol83}, expressing
geometrically the integration by part procedure (a
geometric decomposition was proposed earlier, \eg  by Goldschmidt and
Sternberg  \cite{GoSt73}). Such decomposition formulae, corresponding to the first
and second variational formulae, in particular introduce local
objects such as {\em momenta} which could be globalized by means of
connections. In particular, besides the usual momentum associated with a Lagrangian, a `generalized'
momentum is associated with an Euler-Lagrange type morphism. Its
interpretation in the
calculus of variations has not been exaustively exploited; in \cite{PaVi01} it was
suggested that it could play a r\^ole within the multisymplectic
framework for field theories.

We recall that, by using the so-called interior Euler operator adapted to the finite order case,
a complete representation of the
variational sequence by differential forms was given in
\cite{Kr02,KrMu03,KrMu05} and independently in
\cite{KrSe05,KUV13,VoUr14}. 
This is an operator involved with the integration by parts
procedure. We shall exploit  the relation between the
interior  Euler operator and the Cartan formula for the
Lie derivative of differential forms. The representation of the variational Lie derivative provides, in a quite simple and immediate way, the Noether Theorems as `quotient Cartan formulae' \cite{PRWM15}.
In this paper, inspired by (and extending) a formalism developed by \cite{Kr02,KrMu05}, we shall derive the variational counterpart of the Cartan formula for the Lie derivative of forms of degree $q \leq n-1$, where $n$ is the dimension of the basis manifold, thus explicating and making more precise previous results stated in \cite{FrPaVi02,PRWM15}. In particular formulae for the Lie derivative of classes of $q$-forms will be obtained, not only for the representations.
To this aim we define an interior Euler operator associated with a contact component of degree $k$ of a form of degree $q \leq n-1$.
We shall also define a `momentum' associated with the differential of such a $q$-form and provide an example of application.

The above general results concerning variational derivatives of forms of degree $q \leq n-1$ were motivated by the need to have at hand suitable techniques in order to investigate variational problems for (conserved) currents associated to symmetries and invariant variational problems in locally Lagrangian field theories.
As it is well known, invariance properties of field dynamics are an effective tool to understand a physical system without solving the equations themselves: the existence of conservation laws associated with symmetries of equations strongly simplifies their study and corresponding conserved currents along solutions (on-shell) appear to be significant for the description of the system. 

It turns out to be fundamental to understand whether conserved currents associated with invariance of equations could be identified with Noether conserved currents for a certain Lagrangian;  in fact, a symmetry of a Lagrangian is also a symmetry of its Euler-Lagrange form, but the converse in general is not true. We are interested to this converse problem which belongs to aspects of inverse problems in the calculus of variations.  

%-----------------------------------------------------------------------
\section{Representation of the variational sequence}\label{Noether}
%-----------------------------------------------------------------------

We assume the $r$-th order prolongation of a fibered manifold $\pi : \bY \to \bX$, with $\dim \bX = n$ and $\dim \bY = n+m$, to be the configuration space; this means that {\em fields are assumed to be (local) sections} of $\pi^r : J^r \bY \to \bX$.
We refer to the geometric formulation of the calculus of variations as a subsequence of the de Rham sequence of differential forms on finite order prologations of fibered manifolds.

Due to the affine bundle structure of  
$\pi^{r+1}_{r} : J^{r+1} \bY \to J^{r} \bY$, we have a natural splitting
$J^r\bY\times_{J^{r-1}\bY}T^*J^{r-1}\bY =
J^r\bY\times_{J^{r-1}\bY}(T^*\bX \oplus V^*J^{r-1}\bY)$, which induces natural spittings in  horizontal and vertical parts of vector fields, forms and of the exterior differential on $J^r\bY$ (see the Appendix for some more technical details and properties which will be used here). 

Let $\rho$ be a $q$-form on $J^r\bY$; in particular we obtain a natural decomposition of the pull-back by the affine projections of $\rho$, as 
\beq
(\pi^{r+1}_{r})^*\rho= \sum_{i=0}^{q}p_i\rho \,,
\eeq 
where $p_i\rho$ is the $i$-contact component of $\rho$ (by definition a contact form is zero along any holonomic section of $J^r\bY$).

Starting from this splitting one can define sheaves of contact forms $\Thd^{*}_{r}$, suitably characterized by the kernel of $p_i$ \cite{Kru90}; the sheaves $\Thd^{*}_{r}$ form an exact subsequence of the de Rham sequence on $J^r\bY$ and one can define the quotient sequence
\beq
 0\to \R_{\bY} \to \dots
\to^{\cE_{n-1}} \For^{n}_r/\Thd^{n}_r \to ^{\cE_{n}} \For^{n+1}_r/\Thd^{n+1}_r  \to^{\cE_{n+1}}
\For^{n+2}_r/\Thd^{n+2}_r
\to^{\cE_{n+2}} \dots\to 0 \,
\eeq
\ie the $r$--th order {\em variational sequence} on $\bY\to\bX$ which is an acyclic resolution of the constant sheaf $\R_{\bY}$; see \cite{Kru90}.
In the following, if $\rho\in \For^{k}_r$, then $[\rho]\in \Var^{k}_r \doteq \For^{k}_r /\Thd^{k}_r$ denotes the equivalence class of $\rho$ modulo contact forms as defined by Krupka 
(by an abuse of notation we therefore denote in this way a local or a global section of the sheaf, when there is no possibility of misunderstanding).

The quotient sheaves in the variational sequence can be represented as sheaves of $q$-forms on jet spaces of higher order. 
For $1\leq q \leq n$, the representation mapping is just given by the horizontalization $p_0\rho= h\rho$. For $q > n$, say, $q = n+k$, it is clear that any form is contact; therefore, in this case,  $p_k\rho$ denotes the component of $\rho$ with the lowest degree of contactness.
For $q \geq n+1$, a representation can be given by the {\em interior Euler operator} $\cI$ which is  uniquely intrinsically defined by the decomposition
\beq
p_k\rho=\cI(\rho)+p_kdp_k\cR(\rho) \,,
\eeq 
(where $\cR(\rho)$ is a local $(q-1)$-form) together with the properties 
\beq 
(\pi^{2r+1}_{r})^*\rho -\cI(\rho)\in \Thd^{n+k}_{2r+1}\,\quad \quad 
\cI(p_kdp_k\cR(\rho)) =0 \,; 
\\  
\cI^2(\rho) = (\pi^{4r+3}_{2r+1})^*\cI(\rho)\,\quad\quad
\textstyle{ker} \cI \doteq \Thd^{n+k}_{r}\,.
\eeq
It is defined a representation mapping $R_q: \Var^q_r\to\For^q_s$,
$: [\rho] \mto R_q([\rho])$, with 
\begin{itemize}
\item $R_q([\rho])\doteq p_0 \rho\equiv h\rho$ for $0\leq q\leq n, \, s=r+1$;
\item $R_q([\rho])\doteq \cI(\rho)$ for $n+1\leq q\leq P, \, s=2r+1$;
\item $R_q([\rho])\doteq \rho$ for $P+1\leq q\leq N, \, s=r$;
\end{itemize}
where $N=\textstyle{dim} \,J^r\bY$ and $P$ is the maximal degree of non trivial contact forms on $J^r\bY$ (see \eg \cite{Kr02,KrMu05,Kru90,KrSe05,VoUr14}, whereby also local coordinate expressions can be found). 

The representation sequence $\{ 0\to R_{*}(\Var_{r}^{*})\,,E_{*}\}$, is also exact and we have $E_q\, \com R_q ([\rho])=R_{q+1}\com \, \cE_q ([\rho])= R_{q+1} ([d\rho])$. Currents are sheaf sections $\eps$ of $\Var^{n-1}_{r}$ and $\cE_{n-1}=d_H$ is the total divergence; Lagrangians are sections $\lam$ of $\Var^{n}_{r}$, while $\cE_n$ is called the Euler-Lagrange morphism;
sections $\eta$ of $\Var^{n+1}_{r}$ are called {\em source forms} or also {\em dynamical forms}, while 
$\cE_{n+1}$ is called the Helmholtz morphism. 

It is well known that in order to obtain a representation of Euler-Lagrange type forms the following  integration formula is used \cite{Kru90}
\beq
\ome^{\alp}_{i_{1}   i_{r}  j}\wed \ome_0 = - \, d\ome^{\alp}_{i_{1} i_{r} }\wed \ome_j \,.
\eeq
and the corresponding representation is obtained by taking the $p_1$ component obtained by iterated integrations by parts; here $\ome^{\alp}_{\bI} = d y^{\alp}_{\bI} - y^{\alp}_{\bI j} \wed d x^j $, with $\bI$ a multindex of lenght $r$, are local contact $1$-forms on $J^r \bY$, while we denote by $\ome_0$ the volume density on $\bX$ and by
$\ome_i = \frac{\partial}{\partial x^i} \rfloor \ome_0$, $
\ome_{ij} = \frac{\partial}{\partial x^j} \rfloor \ome_i$ and so on.
In order to integrate by parts $k$-contact components of  $(p+k)$-forms with $p<n$, we notice that
\beq% \label{integration of p+k forms} 
\gam_{\alp}^{i_{1}   i_{r-1}  [i_{r}  j] } \ome^{\alp}_{i_{1}   i_{r-1} [ i_{r} }\wed \ome_{j]} = - \gam_{\alp}^{i_{1}   i_{r-1}  [i_{r}  j] } d \ome^{\alp}_{i_{1}   i_{r-1} }\wed \ome_{i_{r} j} \,.
\eeq
This enabled one of us
%(M.P.)
 to generalize results given in \cite{KrMu05} as follows (see \cite{PRWM15}).

\bPr \label{mathfrak}
Let $\rho \in \For_r^{p+k}$, $1 \leq p \leq n$, $k\geq 1$. 
Let $p_{k} \rho=\sum^{r}_{|\bJ |=0} \ome^{\alp}_{\bJ} \wed \eta^{\bJ}_\alp$, with $\eta^{\bJ}_\alp$ $(k-1)$-contact $(p+k-1)$-forms.
Then we have the decomposition 
\bEq
p_k\rho= \mathfrak{I}(\rho)+p_{k}dp_{k} \mathfrak{R}(\rho) \,,
\eEq
where $\mathfrak{R}(\rho)$ is a local $k$-contact $(p+k-1)$-form such that
\beq
 \mathfrak{I}(\rho)+p_{k}dp_{k} \mathfrak{R}(\rho)  = \ome^{\alp}\wed \sum^{r}_{ |\bJ |= 0}(-1)^{| \bJ |}d_{\bJ} \eta^{\bJ}_\alp+ \sum^{r}_{|\bI |=1}d_{\bI}(\ome^{\alp}\wed \zet^{\bI}_{\alp})\,,
\eeq
with  
$%\beq
\zet^{\bI}_{\alp} = \sum^{r- | \bI |}_{|\bJ |=0}(-1)^{\bJ} \binom{| \bI |+ |\bJ |}{|\bJ |} 
d_{\bJ} \eta^{\bJ \bI}_{\alp}\,.
$%\eeq
\ePr
Note that
$d_{\bJ} \eta^{\bJ}_\alp$ are also $(k-1)$-contact $p$-horizontal $(p+k-1)$-forms.
%It is therefore well defined a local splitting of 
%$p_k\rho$, $\rho \in \For^r_{p+k}$, $1 \leq p \leq n$ and $k > 1$. 
Of course, $\mathfrak{I}=\cI$ and $\mathfrak{R}=\cR$ in the case $p= n$.

\bRm\label{momentum} 
In the case $p=n-1$,  $\mathfrak{R}(\rho)$ is defined  by 
\beq
p_{k}dp_{k}\mathfrak{R}(\rho) = \sum^{r}_{|\bI |=1}d_{\bI}(\ome^{\alp}\wed \zet^{\bI}_{\alp})
= d_H [\sum^{r-1}_{| I |=0}(-1)^k d_{I}\chi^{I [lj]}\wed \ome_{lj}]\,.\eeq
\eRm

%----------------------------------------------
\subsection{Variational Lie derivatives of classes} 
%----------------------------------------------

A {\em variational Lie derivative} operator $\cL_{j^{r}\Xi}$ acting on the sections of sheaves in the variational sequence is well defined: the basic idea is to factorize modulo contact structures \cite{Kru08,KrKrSa10,KrSe05,olga09}. This enables us to define {\em symmetries of classes of forms of any degree} in the variational sequence and corresponding  conservation theorems; see also \cite{PRWM15}.

We define the interior product  of a projectable vector  field
with the equivalence class of $\rho$ as the
equivalence class of the interior product of the vector field with
the representation of the equivalence class of $\rho$, that is
\beq
\iot_{j^r\Xi} [\ome] \equiv j^r\Xi \hat{\rfloor} [\ome] \doteq [ j^s\Xi \rfloor R_q[\ome]] \,.
\eeq
This definition is well given. In fact, we need only to check that the image of $\iot_{j^r\Xi}$ does not change while changing the representative inside the equivalence class. If $[\ome]=[\ome ']$, then $R_q([\ome]-[\ome ']) =0$ (by linearity), therefore $j^r\Xi \hat{\rfloor} ([\ome - \ome ']) $ $=$  $[ j^s\Xi \rfloor R_q[\ome - \ome'] ] =  [0]$ and $j^r\Xi \hat{\rfloor} [\ome] = j^r\Xi \hat{\rfloor} [\ome ']$, as we wanted. 

Accordingly, in the following we will sometimes skip to specify the jet prolongation of a projectable vector field when it appears within formulae for the variational classes (the order in the variational sequence is fixed and so is for the jet order of prolongations).

Therefore we can define the variational Lie derivative with respect to a projectable vector field $(\Xi, \xi)$ of a class of forms in $\Var^q_r$ simply by taking the class of the Lie derivative of its representative with respect to the $s$-prolongation of $\Xi$, \ie 
\beq
\mathcal{L}_{\Xi}   ([\ome]   ) \doteq   [ L_{j^s \Xi} R_q [\ome]   ]\,.
\eeq
As before we can easily check that this definition is well given.

Let $\Xi$ be a projectable vector field on $\bY$, $\rho$ a  $q$-form
defined (locally) on $J^r\bY$. We define an operator 
$\hat{R}_q: \Var^q_r \to \For^q_{s}$ by the following commutativity requirement
\beq
\hat{R}_q \circ \mathcal{L}_{\Xi} =L_{j^{s}\Xi} \circ R_q\,,
\eeq 
\ie $\hat{R}_q \, \mathcal{L}_{\Xi}\, [\rho] = \hat{R}_q\, [ L_{j^{s}\Xi}\, R_q\, [\rho] ]$.
This
operator is uniquely defined and is equal, respectively, to the
following expressions:
\begin{itemize}
\item $L_{j^s\Xi}h\rho \qquad 0\leq q\leq n, \quad s=r+1$;
\item $L_{j^s\Xi} \cI(\rho) \qquad n+1\leq q\leq P, \quad s=2r+1$;
\item $L_{j^s\Xi} \rho \qquad P+1\leq q\leq N, \quad s=r$.
\end{itemize}

This means that $\hat{R}$ together with the (variational) operator $\iot_{\Xi} \equiv \iot_{j^r\Xi}$ return the (differential) operator $i_{j^s \Xi}$, \ie
\beq
\hat{R}_{q-1} \, \iot_{\Xi} [\rho] = i_{j^s \Xi} R_q\, [\rho] = \hat{R}_{q-1} \, [i_{j^s \Xi} R_q[\rho] ]  \,;
\eeq
in the same way, we have that $\hat{R}_{q+1} \circ \mathcal{E}_q = d \circ \hat{R}_{q}$.

This enables us to deal with ordinary Lie derivatives of forms on $\For^q_s$, then  apply the Cartan formula for differential forms, therefore return back to the classes of forms to obtain a sort of {\em variational Cartan formulae}; see in particular also \cite{PRWM15} where the case $q\geq n+1$ has been worked in detail and partial results concerning the case case $q\leq n$ have been obtained.

We shall also need the following naturality property.
\bPr \label{commutativita_e_L}
We have 
 $\mathcal{E}_q \mathcal{L}_{\Xi} = \mathcal{L}_{\Xi} \mathcal{E}_q$.
\ePr

\bPf
 For every $[\ome] \in \Var^q_r$ we have
\beq
 \mathcal{E}_q   ( \mathcal{L}_{\Xi} [\ome] ) = \mathcal{E}_q   [ L_{j^r\Xi} R_q [\ome] ] =   [ d   ( L_{j^r\Xi} R_q [\ome] )] =  [ L_{j^r\Xi} d R_q [\ome] ] \,,
\eeq
on the other hand, $
 \mathcal{L}_{\Xi} ( \mathcal{E}_q [\ome] ) = \mathcal{L}_{\Xi}([d \ome] ) =   [ L_{j^r\Xi} R_{q+1} [d \ome]] $.
The commutator of $d$ and $R_q$ is contact, hence it vanishes in the quotient.
\ePf

%%%%%%---- KRBEK

In the following we shall make use thoroughly of a technical result due to Krbek \cite{Kr02} (Theorem III.11), which we recall here for the convenience of the reader; see
also \cite{KrMu05}.

\bLm \label{Krbek} 
Let $\Psi$ be a $\pi$-vertical vector field on
$\bY$ and $\rho$ a differential \, $q$-form on $J^r\bY$. Then the
following holds true for $i=1,\ldots , q$ 
\beq 
j^{r+2} \Psi \rfloor p_i d p_i \rho = - p_{i-1}d (j^{r+1}\Psi \rfloor p_i\rho)\,, 
\eeq
and 
\beq 
L_{j^{r+2}\Psi}(\pi^{r+2}_{r+1})^* p_i\rho =
j^{r+2}\Psi\rfloor p_{i+1} d p_i \rho +  p_{i}d (j^{r+1}\Psi \rfloor
p_i\rho)\,, 
\eeq
\eLm

%-------------------------------------------------------------------
\subsection{The case $ q \leq n-1$}
%-------------------------------------------------------------------

\bDf
The momentum associated with the density $\alp = [\rho] \in \Var^q_r $ and the projectable vector field $\Xi$ is defined as a section 
$\tilde{p}_{d_V \alp} \in \Var^q_{r+1}$, of which the  representation 
$R_k (\tilde{p}_{d_V \alp}) = \tilde{p}_{d_V R_k \alp} = \tilde{p}_{d_V h \rho}$ is a local $1$-contact $q$-form satisfying the identity
\beq
d_H ( j^{s-1}\Xi_V {\rfloor} \tilde{p}_{d_V h \rho} ) = - \, d_H ( j^{s-1}\Xi_V \rfloor p_1 \mathfrak{R}(d \rho) )\,,
\eeq
where $\mathfrak{R}$ is defined by the splitting of Proposition \ref{mathfrak}.
\eDf

%-----------------------------------------------------

We have the following.
\bTh \label{Cartan_identity_1}
Let $\alp \in \Var^q_r$, $0 \leq q \leq n-1$, and let
$\Xi$ be a $\pi$-projectable  vector field on $\bY$; the following holds locally
\beq 
 \cL_{\Xi} \alp = \Xi_H \hat{\rfloor} \mathcal{E}_q (\alp)+ \mathcal{E}_{q-1} (\Xi_V \hat{\rfloor} \tilde{p}_{d_V \alp}+\Xi_H \hat{\rfloor} \alp)\,.
\eeq 
\eTh

\bPf
Recalling the decomposition of the vector fields and of the exterior differential we have
\beq
 (\pi^{r+3}_{r+1})^* \hat{R}_q \mathcal{L}_\Xi [\rho] = (\pi^{r+3}_{r+1})^* L_{j^{r+1} \Xi} R_q [\rho] = (\pi^{r+3}_{r+1})^* L_{j^{r+1} \Xi} (h \rho) =
 \eeq 
\beq
 = (\pi^{r+3}_{r+1})^* {j^{r+1} \Xi} \rfloor d (h \rho) + (\pi^{r+3}_{r+1})^* d ({j^{r+1} \Xi} \rfloor h \rho) =
 \eeq 
\beq
  = ({j^{r+1} \Xi}_H + {j^{r+1} \Xi}_V) \rfloor (d_H + d_V) h \rho + (d_H + d_V) ({j^{r+1} \Xi}_H + {j^{r+1} \Xi}_V) \rfloor h \rho \,.
  \eeq
By applying lemmas \ref{ext diff},  \ref{commutativita_pi_p} and $\ref{contrazioni_campi_orizzontali_verticali}$, we easily see that:
\beq
{j^{r+1} \Xi}_V \rfloor (d_H h \rho) = {j^{r+1} \Xi}_V \rfloor h (d h^2 \rho) = 0 \,,
\eeq
it is also easy to check that
$d_H ({j^{r+1} \Xi}_V \rfloor h \rho)$ and
$d_V ({j^{r+1} \Xi}_V \rfloor h \rho)$ vanish, while
\beq
 {j^{r+1} \Xi}_H \rfloor (d_V h \rho) = {j^{r+1} \Xi}_H \rfloor (p_1 d h^2 \rho) \,, \qquad
 d_V ({j^{r+1} \Xi}_H \rfloor h \rho) = p_1 d h ({j^{r+1} \Xi}_H \rfloor h \rho) \,,
\eeq
are contact pieces. On the other hand, by using Lemma $\ref{lemma_krbek}$
\beq
{j^{r+1} \Xi}_H \rfloor d_H (h \rho) = {j^{r+1} \Xi}_H \rfloor d_H R_q [\rho] = 
 \eeq 
\beq
= {j^{r+1} \Xi}_H \rfloor (\pi^{r+3}_{r+1})^* R_{q+1} \mathcal{E}_q [\rho] = 
(\pi^{r+3}_{r+1})^* \hat{R}_q (\Xi_H \rfloor \mathcal{E}_q  [\rho])\,,
\eeq
and
\beq
 d_H ( {j^{r+1} \Xi}_H \rfloor h \rho ) = d_H ({j^{r+1} \Xi}_H \rfloor R_q [\rho] ) = (d_H + d_V) ( {j^{r+1} \Xi}_H \rfloor R_q [\rho] ) = 
 \eeq 
\beq
 = (\pi^{r+3}_{r+1})^* d ({j^{r+1} \Xi}_H \rfloor R_q [\rho] ) = (\pi^{r+3}_{r+1})^* \hat{R}_q [d ({j^{r+1} \Xi}_H \rfloor R_q [\rho] ) ] = 
 \eeq 
\beq
 = (\pi^{r+3}_{r+1})^* \hat{R}_q \mathcal{E}_q [{j^{r+1} \Xi}_H \rfloor R_q [\rho] ] = (\pi^{r+3}_{r+1})^* \hat{R}_q \mathcal{E}_q (\Xi_H \rfloor [\rho] )\,.
\eeq
Analogously, one can see that
\beq
{j^{r+1} \Xi}_V \rfloor d_V h \rho 
%= (\pi^{r+3}_{r+1})^* \hat{R}_q (\Xi_V \rfloor [d_V R_q [\rho]]) 
= (\pi^{r+3}_{r+1})^* \hat{R}_q[ j^{r+1} \Xi_V \rfloor d_V  \rho] \,,
\eeq
 so that up to contact terms
\beq
 (\pi^{r+3}_{r+1})^* \hat{R}_q \mathcal{L}_\Xi [\rho] = {j^{r+1} \Xi}_V \rfloor d_V h \rho + {j^{r+1} \Xi}_H \rfloor d_H (h \rho) + d_H ( {j^{r+1} \Xi}_H \rfloor h \rho ) = 
 \eeq 
\beq
 = (\pi^{r+3}_{r+1})^* \hat{R}_q( [j^{r+1}  \Xi_V \rfloor d_V  \rho] + \Xi_H \rfloor \mathcal{E}_q [\rho] + \mathcal{E}_{q-1} (\Xi_H \rfloor [\rho]) ) \,.
\eeq
By taking the class, which makes the remaining contact pieces vanish, 
\beq
 \mathcal{L}_{\Xi} [\rho] = [ j^{r+1} \Xi_V \rfloor d_V  \rho] + \Xi_H \hat{\rfloor} \mathcal{E}_q ([\rho] )+ \mathcal{E}_{q-1} (\Xi_H \hat{\rfloor} [\rho])\,.
\eeq
By splitting $ d_V \rho$ according with Proposition \ref{mathfrak}, again by  Lemma $\ref{Krbek}$, and since $ [ j^{2r+1} \Xi_V \rfloor \mathfrak{I}(d \rho)] 
= 0 $, 
the result is obtained 
 by denoting $\alp = [\rho]$.
\ePf

%-------------------------------------------------------------------
\subsection{The case $q =  n $}
%-------------------------------------------------------------------

\bDf
The momentum associated with the  Lagrangian $\alp = [\rho] \in \Var^n_r$ and the projectable vector field $\Xi$ is defined as a section $p_{d_V \alp} \in \Var^n_{r+1}$,  of which the representation
$R_k (p_{d_V \alp}) = p_{d_V R_k \alp} = p_{d_V h \rho}$ is a local $1$-contact $n$-form satisfying the identity
\beq
d_H ( \Xi_V \rfloor p_{d_V h \rho} ) = - d_H ( \Xi_V \rfloor p_1 \mathcal{R}(d \rho) )\,,
\eeq
where $\mathcal{R}$ is defined by the splitting given by the interior Euler operator.
\eDf

\bTh\label{Cartan_identity_2}(Noether's Theorem I)

Let $\alp \in \Var^n_r $ and 
$\Xi$ be a $\pi$-projectable  vector field on $\bY$;
the following holds (locally): 
\beq
\mathcal{L}_{\Xi} \alp = \Xi_V \hat{\rfloor} \mathcal{E}_n (\alp) + \mathcal{E}_{n-1} (\Xi_V \hat{\rfloor} p_{d_V \alp} + \Xi_H \hat{\rfloor} \alp)\,.
\eeq
\eTh

\bPf
As before, by the representation $\hat{R}_n$ and the pullback $(\pi^{r+3}_{r+1})^*$ 
\beq
 (\pi^{r+3}_{r+1})^* \hat{R}_n \mathcal{L}_\Xi [\rho] = {j^{r+1} \Xi}_V \rfloor d_V h \rho + {j^{r+1} \Xi}_H \rfloor d_H h \rho + d_H ({j^{r+1} \Xi}_H \rfloor h \rho) \,,
\eeq however (unlike the case $k \leq  n-1 $)
the term ${j^{r+1} \Xi}_H \rfloor d_H h \rho$ vanishes because $d\rho$ is contact and
$d_H h \rho = (\pi^{r+3}_{r+1})^* h (d \rho) = 0$.
On the other hand
\beq
 d_V h \rho = p_1 d h^2 \rho = (\pi^{r+3}_{r+2})^* p_1 d h \rho = 
 \eeq 
\beq
 = (\pi^{r+3}_{r+2})^* [ (\pi^{r+2}_{r+1})^* (p_1 d) - p_1 d p_1 ] \rho =  
 \eeq 
\beq
 = (\pi^{r+3}_{r+1})^* \mathcal{I} (d \rho) + (\pi^{r+3}_{r+1})^* p_1 d p_1 \mathcal{R}(d \rho) - (\pi^{r+3}_{r+2})^* p_1 d p_1 \rho\,.
\eeq
Thus, by Krbek's Lemma we get
\beq
 (\pi^{r+3}_{r+1})^* \hat{R}_n \mathcal{L}_\Xi [\rho] 
= {j^{r+1} \Xi}_V \rfloor (\pi^{r+3}_{r+1})^* \mathcal{I} (d \rho) - h d_H ( {j^{r+1} \Xi}_V \rfloor p_1 \mathcal{R}(d \rho) ) + 
 \eeq 
\beq
+ (\pi^{r+3}_{r+2})^* h d (j^{r+1} \Xi_V \rfloor p_1 \rho) + d_H ({j^{r+1} \Xi}_H \rfloor h \rho)=
 \eeq 
\beq
= {j^{r+1} \Xi}_V \rfloor (\pi^{r+3}_{r+1})^* \mathcal{I} (d \rho) + h d_H ( {j^{r+1} \Xi}_V \rfloor p_{d_V h \rho} ) + 
 \eeq 
\beq
+ (\pi^{r+3}_{r+2})^* h d (j^{r+1} \Xi_V \rfloor p_1 \rho) + d_H ({j^{r+1} \Xi}_H \rfloor h \rho)\,.
\eeq
However, by Lemma $\ref{lemma_krbek}$,
$(\pi^{r+3}_{r+2})^* h d (j^{r+1} \Xi_V \rfloor p_1 \rho) (\bullet)= d_H h p_1 \rho (\Xi_V, \bullet) = 0$.
Lastly, we use again the representations in each remaining term:
\beq
{j^{r+1} \Xi}_V \rfloor (\pi^{r+3}_{r+1})^* \mathcal{I} (d \rho) = {j^{r+1} \Xi}_V \rfloor (\pi^{r+3}_{r+1})^* R_{n+1} [d \rho] = 
 \eeq 
\beq
 = {j^{r+1} \Xi}_V \rfloor (\pi^{r+3}_{r+1})^* R_{n+1} \mathcal{E}_n [\rho] = (\pi^{r+3}_{r+1})^* \hat{R}_n ( \Xi_V \rfloor \mathcal{E}_n [\rho] )\,;
\eeq

\beq
h d_H ({j^{r+1} \Xi}_V \rfloor p_{d_V h \rho}) = (\pi^{r+3}_{r+1})^* h d ({j^{r+1} \Xi}_V \rfloor p_{d_V h \rho}) =
 \eeq 
\beq
 = (\pi^{r+3}_{r+1})^* R_n [ d ({j^{r+1} \Xi}_V \rfloor p_{d_V R_n [\rho]}) ] = (\pi^{r+3}_{r+1})^* R_n \mathcal{E}_{n-1} [{j^{r+1} \Xi}_V \rfloor R_n (p_{d_V [\rho]}) ] = 
 \eeq 
\beq
 = (\pi^{r+3}_{r+1})^* R_n \mathcal{E}_{n-1} ( \Xi_V \hat{\rfloor} p_{d_V [\rho]} ) = (\pi^{r+3}_{r+1})^* \hat{R}_n \mathcal{E}_{n-1} ( \Xi_V \hat{\rfloor} p_{d_V [\rho]} ) \,;
\eeq

\beq
d_H ({j^{r+1} \Xi}_H \rfloor h \rho) = (d_V + d_H) ({j^{r+1} \Xi}_H \rfloor h \rho) = (\pi^{r+3}_{r+1})^* d ({j^{r+1} \Xi}_H \rfloor h \rho) = 
 \eeq 
\beq
 = (\pi^{r+3}_{r+1})^* d ({j^{r+1} \Xi}_H \rfloor h \rho) = (\pi^{r+3}_{r+1})^* \hat{R}_n [ d ( {j^{r+1} \Xi}_H \rfloor h \rho ) ] 
 \eeq 
\beq
 = (\pi^{r+3}_{r+1})^* \hat{R}_n \mathcal{E}_{n-1} [ {j^{r+1} \Xi}_H \rfloor R_n [\rho] ] = (\pi^{r+3}_{r+1})^* \hat{R}_n \,\mathcal{E}_{n-1} (\Xi_H \hat{\rfloor} [\rho] )\,.
\eeq
As before, by calling $\alp = [\rho]$, we get the conclusion.
\ePf

%-------------------------------------------------------------------
\subsection{The case $q \geq n+1 $}
%-------------------------------------------------------------------

In \cite{PRWM15} it was proved the following
 variational Cartan formula for classes of forms of degree $q \geq n+1$ (the case $q =n+1$ for locally variational dynamical forms encompasses Noether's Theorem II, or so-called Bessel-Hagen symmetries).
\bTh\label{HigherLieDer} 
Let $q= n + k$,  with $ k \geq 1$ and $\alp \in \Var^q_r $. Let
$\Xi$ be a $\pi$-projectable  vector field on $\bY$; we have \beq
\cL_{\Xi}\alp =  \Xi_V \hat{\rfloor} \cE_{q}(\alp) +
\cE_{q-1}(\Xi_V \hat{\rfloor} \alp)\,. 
\eeq 
\eTh

These variational Cartan formulae will be the underlying mathematical core of the next Section, which deals with currents associated with invariance of
(locally) variational dynamical forms, invariance of currents and corresponding generalized momenta.

%------------------------------------------------------------
\section{Noether--Bessel-Hagen currents}
%------------------------------------------------------------

Consider now conserved currents associated with invariance properties of {\em (locally) variational} {\em global} field equations, \ie with so-called generalized or Bessel-Hagen symmetries \cite{BeHa21}.
Noether currents for different local Lagrangian presentations and corresponding {\em conserved currents associated with each local presentation}  have been characterized in 
\cite{FePaWi10,FrPaWi12,FrPaWi13,PaWiGa12}.
There exist cohomological obstructions for such local currents be globalized
and such obstructions are also related with the {\em existence of global solutions} for a given global field equation \cite{FrPaWi13b}.

We will denote by a subscript $i$ the fact that in general a sheaf section is defined only locally, \ie that it is a $0$-cochain in \v{C}ech cohomology; analogously by two subscripts $ij$ we shall denote that a sheaf section is a $1$-cochain.
In the following we shall also denote  $\hat{\rfloor}$ simply by $\rfloor$ since we are dealing with classes and there is no danger of confusion.
Let for simplicity $\eta_{\lam_i}$ denote a global Euler--Lagrange class of forms for a 
(local) variational problem represented by (local) sheaf sections $\lam_i$. 
Notice that in this case Theorem \ref{Cartan_identity_2} (Noether Theorem I) reads
$\cL_{\Xi} \lam_{i}= \Xi_V\rfloor\eta_{\lam_i} +d_H \eps_i$;
where $\eps_i = \Xi_V\rfloor p_{d_V\lam_i} +\Xi_H \rfloor \lam_i$ is the Noether current associated with it.
\bDf
A generalized symmetry of a (locally variational) dynamical form $\eta_{\lam_i}$ is
a projectable vector field $j^r\Xi$ on $J^r\bY$ such that $\cL_{\Xi}\eta_{\lam_i}= 0$.
\eDf
Since we assume $\eta_{\lam_i}$ to be closed, Theorem \ref{HigherLieDer} reduces (case $q=n+1$) to $\cL_{\Xi} \eta_{\lam_i} =\cE_n (\Xi_V\rfloor \eta_{\lam_i})$, and  if $j^r\Xi$ is such that $\cL_{\Xi} \eta_{\lam_i} =0$, then $\cE_n (\Xi_V\rfloor \eta_{\lam_i}) = 0$;
therefore, {\em locally}  we have
$\Xi_V\rfloor\eta_{\lam_i} =d_H\nu_i$.
Notice that, although $\Xi_V\rfloor \eta_{\lam_i}$ is global, in general it defines a non trivial cohomology class \cite{FePaWi10}; it is clear that $\nu_i$ is a (local) current which is conserved on-shell (\ie along critical sections).
On the other hand, {\em and  independently} (see \cite{Noe18}), we get {\em locally}
$\cL_{\Xi} \lam_{i} = d_H \beta_{i}$ thus  we can write
$\Xi_{V} \rfloor \eta_{\lam}  + d_{H}( \eps_i  -  \beta_{i} )$  $=$ $0$, where $\eps_i$ is the usual {\em canonical} Noether current. 
\bDf
We call the (local) current $\eps_{i} - \beta_{i}$ a {\em Noether--Bessel-Hagen current}.
\eDf
A Noether--Bessel-Hagen current  $\eps_{i} - \beta_{i}$ is a current (conserved along critical sections) associated with a generalized symmetry; in \cite{FrPaWi13,FrPaWi13b} we proved that a Noether--Bessel-Hagen current is variationally equivalent to a global (conserved) current if and only if $0 = [[\Xi_{V} \rfloor \cE_{n}(\lam_{i})]] \in H^{n}_{dR}(\bY)$.

According with our general considerations, in view of the precise statements which the Noether Teorems provide concerning the existence and the nature of conservation laws for invariant variational problems,  it is of importance to determine whether a Noether--Bessel-Hagen current is variationally equivalent to a Noether conserved current for a suitable invariant Lagrangian.
It is known that this is involved with the existence of a variationally trivial local Lagrangian $d_H\mu_i$, and with a condition on the current associated with it \cite{PaWi14}.
In the following we will relax some of the conditions and investigate the outcome.

\bPr\label{NBH}
A Noether--Bessel-Hagen current $\eps_{\lam_i} -\bet_i$ associated with a generalized symmetry of $\eta_{\lam_i}$ is a Noether conserved current (for that symmetry) if and only if it is of the form $\eps_{\lam_i}  - \cL_{\Xi}\mu_i$, with 
$\mu_i$ a current satisfying $\cL_{\Xi}(\lam_i - d_H\mu_i)=0$.
\ePr

\bPf
From $\cL_{\Xi}\eta_{\lam_i}=0$, we get $\cL_{\Xi}\lam_i = d_H\bet_i$.  It is easy to see that the current $\eps_{\lam_i} -\bet_i$ is a Noether conserved current if and only if there exists $\mu_i$ such that $\bet_i - \cL_{\Xi}\mu_i 
%%%%%%%%%%%%%%%%%%%%%%%
$ is closed, \ie  if and only if 
\beq
d_H\bet_i = d_H(\Xi_V\rfloor p_{d_V d_H \mu_i} +\Xi_H\rfloor d_H\mu_i) \,.
\eeq
 This means of course, that,  locally, $\bet_i = \cL_{\Xi}\mu_i 
%%%%%%%%%%%%%%%%%%%%%%%
+d_H\gam_{ij} $. 
On the other hand $d_H\bet_i = d_H \cL_{\Xi}\mu_i = d_H(\Xi_H\rfloor d_H\mu_i)$. Notice that, comparing the two expression we get $d_H({\Xi_V} \rfloor p_{d_V d_H \mu_i})\equiv 0$; thus, in particular,  this identity is a consequence of the fact that $\cL_{\Xi}$ commutes with $d_H$.
\ePf

\bPr
Let $\bet_i = \cL_{\Xi}\mu_i$ (\ie $d_H \gam_{ij} =0$).
The Noether current $\eps_{\lam_i - d_H \mu_i}$ is exact on-shell and it is equal to $d_H (\Xi_V \rfloor \tilde{p}_{d_V \mu_i}+\Xi_H \rfloor \mu_i)$.
\ePr

\bPf
 As it is well known, along any section pulling back to zero $\Xi_V \rfloor \eta_{\lam_i}$ we get the on-shell conservation law $d_H(\eps_{\lam_i} -\bet_i )=0$.  If there exists a current $\mu_i$ such that $\bet_i = \cL_{\Xi}\mu_i \equiv \Xi_H\rfloor d_H\mu_i+ d_H(\Xi_V\rfloor \tilde{p}_{d_V\mu}+\Xi_H\rfloor \mu_i)$, therefore
$\Xi_V\rfloor p_{d_V\lam_i}+\Xi_H\rfloor (\lam_i - d_H\mu_i) - d_H(\Xi_V\rfloor \tilde{p}_{d_V\mu_i}+\Xi_H\rfloor \mu_i)$ 
is closed on-shell.
By a uniqueness argument,  we see that the latter expression must be equal to $\Xi_V\rfloor p_{d_V d_H \mu_i}$; therefore $d_H(\Xi_V\rfloor \tilde{p}_{d_V\mu_i}+\Xi_H\rfloor \mu_i)= \eps_{\lam_i - d_H\mu_i}$ on-shell. 
\ePf

\bRm  
It turns out that, on-shell, a canonical potential of the Noether current  
$\eps_{\lam_i - d_H\mu_i}$, then a corresponding conserved quantity, is defined.
An {\em off-shell} \,exact Noether current associated with the invariance of $\lam_i-d_H\mu_i$ would be generated by a generalized symmetry $j^r\Xi$ such that $\Xi_V\rfloor\eta_{\lam_i}=0$;  the corresponding cohomology class would be, therefore, trivial (see the discussion in \cite{FrPaWi13,FrPaWi13b}). 
\eRm

Our next goal is to relax the results above by recasting the problem by using directly the definition of canonical Noether current rather than the splittings of the Lie derivative given by the Cartan identities.
So, it could be useful to consider $\eps_i$ not just as a single current but as a morphism of the type $\eps_i: \lam_i \mapsto \eps_{\lam_i}$, from the sheaf of the Lagrangians to the one of the currents. It is obviously linear since the interior product, the vertical differential and the momentum are so.

It is now instructive to obtain the  result of Proposition \ref{NBH} as a consequence of  Krbek's Lemma.

First we state some preliminary technical results.
\bLm
We have 
\beq
d_H (\Xi_V \rfloor d_V \mu_i) = d_H ( \Xi_V \rfloor \mathfrak{I}(d \mu_i) )\,.
\eeq
\eLm

\bPf 
Since, up to pullback, $hd = d_H h = h d_H$, we have
\beq
\Xi_V \rfloor d_V \mu_i = \Xi_V \rfloor \mathfrak{I}(d \mu_i) +\Xi_V \rfloor p_1 d p_1 \mathfrak{R}(d \mu_i) = \Xi_V \rfloor \mathfrak{I}(d \mu_i) - h d (\Xi_V \rfloor p_1 \mathfrak{R}(d \mu_i) ) =
 \eeq 
\beq
 = \Xi_V \rfloor \mathfrak{I}(d \mu_i) - h d_H (\Xi_V \rfloor p_1 \mathfrak{R}(d \mu_i) ) = \Xi_V \rfloor \mathfrak{I}(d \mu_i) - d_H ( \Xi_V \rfloor \tilde{p}_{d_V \mu_i} ) \,.
\eeq
The statement follows immediately.
\ePf

\bLm
We have
\beq
  \Xi_V \rfloor p_{d_V (d_H \mu_i)} = \Xi_V \rfloor \mathfrak{I}(d \mu_i)+ d_H(\Xi_V \rfloor \tilde{p}_{d_V \mu_i} + \Xi_H \rfloor \mu_i)\,.
\eeq
\eLm

\bPf
It is a consequence of the naturality of the variational Lie derivative: $\mathcal{L}_\Xi (d_H \mu_i) = d_H ( \mathcal{L}_\Xi \mu_i ) $.
In fact, from one side
\beq
\mathcal{L}_\Xi (d_H \mu_i) = d_H ( \Xi_V \rfloor p_{d_V (d_H \mu)} +
 \Xi_H \rfloor d_H \mu_i ) = 
d_H ( \Xi_V \rfloor p_{d_V (d_H \mu_i)}) + d_H \bet_i \,.
\eeq
while on the other side
\beq 
d_H ( \mathcal{L}_\Xi \mu_i ) = d_H ( \Xi_V \rfloor d_V \mu_i + \Xi_H \rfloor d_H \mu_i + d_H (\Xi_H \rfloor \mu_i ) ) = d_H ( \Xi_V \rfloor d_V \mu_i ) + d_H \bet_i
\,;
\eeq
hence $d_H ( \Xi_V \rfloor d_V \mu_i ) = d_H ( \Xi_V \rfloor p_{d_V (d_{H} \mu_i)} )$. Therefore, by the formula proved above, $d_H ( \Xi_V \rfloor \mathfrak{I}(d \mu_i) ) = d_H ( \Xi_V \rfloor p_{d_V (d_{H} \mu_i )} )$ as well, and $\Xi_V \rfloor \mathfrak{I}(d \mu _i) = \Xi_V \rfloor p_{d_V (d_{H} \mu_i )} + d_{H} \phi_{ij}$. We therefore  get the result.
\ePf

\bLm %\label{lemma_mio}%Cattafi
 Given $\mu_i \in \Var^{n-1}_r$, we have
\beq
d_H ( \Xi_V \rfloor p_{d_V d_H \mu_i } ) = 0\,.
\eeq
\eLm

\bPf
From one side we have:
\beq (\pi^{r+5}_{r+3})^*d_V ( R_n d_H \mu_i ) = d_V ( (\pi^{r+3}_{r+1})^* R_n d_H \mu_i ) = d_V ( d_H R_{n-1} \mu_i ) = d_V ( d_H h \mu_i ) =  
 \eeq 
\beq
= d_V (h d_H \mu_i)= d_V ( (h d_H + h d_V) \mu_i ) = d_V h (\pi^{r+2}_r)^*d \mu_i = (\pi^{r+5}_{r+3})^*d_V h d \mu_i\,.
\eeq
On the other hand, since the pull--back $(\pi^{r+5}_{r+3})^*$ is injective, we have by definition
\beq R_n d_H ( \Xi_V \rfloor p_{d_V d_H \mu_i } ) = d_H R_{n-1} ( \Xi_V \rfloor p_{d_V d_H \mu_i } ) = d_H ( \Xi_V \rfloor R_n ( p_{d_V d_H \mu_i } ) ) = 
 \eeq 
\beq
 = d_H ( \Xi_V \rfloor p_{d_V R_n ( d_H \mu_i ) } ) = d_H ( \Xi_V \rfloor p_{d_V h (d \mu_i) } ) = - d_H ( \Xi_V \rfloor p_1 \mathcal{R}(d (d \mu_i  ) ) = 
 0\,,
  \eeq
which gives us the assertion.
\ePf

Let us suppose now that
$\bet_i$ is a Noether current associated with the Lagrangian $\lam_i - \alp_i$, with $\alp_i =d_H \mu_i$,
\ie $\bet_i = \eps_{\lam_i -d_H \mu_i}$.

\bPr
The Noether-Bessel-Hagen current $\eps_{\lam_i } - \bet_i$ is the canonical Noether current associated with the Lagrangian $\lam_i - d_H \mu_i$ if and only if $\bet_i = \Xi_H \rfloor d_H \mu_i$ modulo a locally exact current.
\ePr

\bPf
First of all, by linearity, we see that $\eps_{\lam_i} - \bet_i$ is a Noether current associated with $\lam_i - d_H \mu_i$ if and only if $\eps_{\lam_i} - \bet_i= \eps_{\lam_i - d_H \mu_i} = \eps_{\lam_i}- \eps_{d_H \mu_i}$, \ie if and only if $\bet_i= \eps_{d_H \mu_i}$.

On the other hand, the previous 
Lemma implies that $\Xi_V \rfloor p_{d_V d_H \mu_i}$ is closed, hence locally exact ($\Xi_V \rfloor p_{d_V d_H \mu_i } = d_H \gam_{ij}$). Thus, by definition of Noether current associated to $d_H \mu_i$,
\beq
 \eps_{d_H \mu_i } = \Xi_V \rfloor p_{d_{V} d_{H} \mu_i } + \Xi_H \rfloor d_H \mu_i = d_H \gam_{ij} + \Xi_H \rfloor d_H \mu_i \,.
 \eeq

This means that $\bet_i$ is the Noether current associated with $\lam_i - d_H \mu_i$ if and only if $\bet_i= \Xi_H \rfloor d_H \mu_i + d_H \gam_{ij}$. 

As we saw, the ``if'' implication could be weakened further since
\beq
 \mathcal{L}_{\Xi} ( d_H \mu_i ) = %\Xi_V \rfloor \cancel{ \mathcal{E}_n d_H \mu} + 
\mathcal{E}_{n-1} ( \Xi_V \rfloor p_{d_V d_H \mu_i} + \Xi_H \rfloor d_H \mu_i ) 
=  \mathcal{E}_n ( \Xi_H \rfloor d_H \mu_i ) \,,
\eeq
and, by the uniqueness of the decomposition of the Lie derivative, the condition $\bet_i = \Xi_H \rfloor d_H \mu_i $ is sufficient in order to have $\bet_i= \eps_{d_H \mu_i }$, \ie we can take $d_H\gam_{ij}= 0$.
Conversely, the indetermination remains because, when $\bet_i= \eps_{d_H \mu_i }$, only the differential of $\bet_i$ and $\Xi_H \rfloor d_H \mu_i$ are equal.
\ePf
However, we can still state the following relaxed result.

\bPr
Under the hypothesis $\bet_i = \Xi_H \rfloor d_H \mu_i + d_H \gam_{ij}$, the Noether-Bessel-Hagen current $\eps_{\lam_i} - \bet_i= \eps_{\lam_i - d_H \mu_i}$ is  exact on-shell, and its potential $\Xi_V \rfloor \tilde{p}_{d_V \mu_i} + \Xi_H \rfloor \mu_i$  is defined up to a cohomology class.
\ePr

\bPf
The on shell conservation law $d_H (\eps_{\lam_i} -\bet_i )=0$ implies 
\beq
 \Xi_H \rfloor d_H \mu_i + d_H ( \Xi_V \rfloor \tilde{p}_{d_V \mu_i} + \Xi_H \rfloor \mu_i )+ \Xi_V \rfloor \mathfrak{I}(d \mu_i ) = \Xi_V \rfloor p_{d_V \lam_i } + \Xi_H \rfloor \lam_i +d_H \psi_{ij}\,;
\eeq
by simple manipulations, thanks to the Lemmas above, we obtain 
\beq
\eps_{\lam_i -d_H\mu_i} =  d_H(\Xi_V \rfloor \tilde{p}_{d_V \mu_i } + \Xi_H \rfloor \mu_i 
+d_H \tilde{\psi_{ij}}) \,. 
\eeq
\ePf
Note that  (unlike the case $d_H\gam_{ij} =0$), generally speaking, the potential of the Noether--Bessel-Hagen current is not a canonical one. 

\subsection*{Acknowledgement}

Research supported by Department of Mathematics-University of Torino research project {\em Geometric methods in mathematical physics and applications} $2013-14$ (M.P.) and $2014-15$ (E.W.); F.C. was partially supported by the NWO VIDI project {\em Poisson Geometry Inside Out} $639.033.312$.

\section{Appendix}
%%%%%%%%% da Tesi cattafi parte introduttiva

For the convenience of the reader, we recall some useful technical tools needed in Section \ref{Noether}; details can be found \eg in \cite{Cat15,Kr02,Sau89}.

\bLm \label{contrazioni_campi_orizzontali_verticali}
 Given the vector field $X$ and the differential form $\rho$, the contraction between $X_V$ and the horizontal component $h \rho$ is zero; the same holds for the contraction between $X_H$ and the $n$-contact component $p_n \rho$.
\eLm
\bLm \label{commutativita_pi_p}
 For every $\rho \in \Lam^k(J^rY)$, $p_i^2 \rho = (\pi^{r+2}_{r+1})^* (p_i \rho) = p_i (\pi^{r+1}_r)^* \rho $ $ \forall i$.
\eLm

\bPf
Since  for every $\rho \in \Lam^k(J^rY)$, $p_j p_i \rho = 0$ $\forall i \neq j$, it is enough to apply the decomposition formula twice, first on $p_i \rho$, then on $\rho$:
\beq 
(\pi^{r+2}_{r+1})^* (p_i \rho) = \sum_{j=1}^k p_j (p_i \rho) = p_i^2 \rho = p_i (p_i \rho) = p_i \sum_{j=1}^k p_j \rho = p_i (\pi^{r+1}_r)^* \rho \,.
\eeq
In particular, the operators $p_i$ behave almost like projectors: their composition is not the Kronecker delta, but we have the following formula
\beq
p_i p_j = \del_{ij} (\pi^{r+2}_{r+1})^* p_j = \del_{ij} p_j (\pi^{r+1}_r)^*\,.
\eeq
\ePf
\bLm \label{ext diff}
We have the following decomposition of the exterior differential:
\beq
(\pi^{r+2}_r)^*d = d_H + d_V \,.
\eeq
\eLm

\bPf
Thanks to the contravariance of the pullback,
\beq
(\pi^{r+2}_r)^* (d \rho) = (\pi^{r+1}_r \circ \pi^{r+2}_{r+1} )^* (d \rho) = (\pi^{r+2}_{r+1})^*   ( (\pi^{r+1}_r)^* (d \rho)   ) = 
 \eeq 
\beq
 = (\pi^{r+2}_{r+1})^* \sum_{i=0}^k p_i (d \rho) = \sum_{i=0}^k (\pi^{r+2}_{r+1})^* p_i (d \rho) = \sum_{i=0}^k (p_i d p_{i-1} \rho + p_i d p_i \rho) =
 \eeq 
\beq
  = \sum_{i=0}^k (p_i d p_{i-1} \rho) + \sum_{i=0}^k (p_i d p_i \rho) = \sum_{i=0}^{k-1} (p_{i+1} d p_i \rho) + \sum_{i=0}^k (p_i d p_i \rho) = d_V \rho + d_H \rho\,.
  \eeq
\ePf

We have several fundamental properties, which relates the operators $p_i$, $h$, $d$, $d_H$ and $d_V$
\bLm\label{lemma_krbek}
 For every $i \geq 1$, supposing the operators are applied to $k$-forms,
\begin{enumerate}
 \item $p_i d_H = d_H p_i$
 \item $p_i d_V = d_V p_{i-1}$
 \item $(\pi^{r+2}_{r+1})^*(p_i d) = p_i d (p_i + p_{i-1})$

 \item $h d_H = d_H h$
 \item $h d_V = 0$
 \item $(\pi^{r+3}_{r+1})^* (h d) = d_H h = h d_H$

 \item $d_H^2 = 0$
 \item $d_V^2 = 0$
 \item $d_H d_V = -d_V d_H$\,.
\end{enumerate}
\eLm

%-------------------------------------

\end{document}